\documentstyle[emulateapj]{article}
 
\def\al{\alpha}
\def\cd{{\cal{D}}}

\def\dt{\Delta t_{\rm obs}}
\def\eps{\epsilon}
\def\epsg{\epsilon_{\gamma}}
\def\epsp{{\epsilon^{\prime}}}
\def\epspp{{\epsilon^{\prime \prime}}}
\def\gam{\gamma}
\def\Gam{\Gamma}

\def\lssc{{l_{\rm ssc}}}
\def\lrsc{{l_{\rm rsc}}}

\def\npp{{n^{\prime \prime}}}

\def\nursc{{\nu_{\rm rsc}}}
\def\nusp{{\nu_{\rm s}^{\prime}}}
\def\rblr{{R_{_{\rm BLR}}}}
\def\siggg{\sigma_{\gamma \gamma}}
\def\st{\sigma_{_{\rm T}}}
\def\tblr{\tau_{_{\rm BLR}}}
\def\tgg{\tau_{\gamma \gamma}}
\def\tggrsc{\tgg^{\rm rsc}(\eps_{\rm obs})}

\def\usynp{u_{\rm syn}^{\prime}}
\def\veps{\varepsilon}

\slugcomment{To Appear in The Astrophysical Journal 538, No1. 20 July Issue, 
2000}

\begin{document}
       
\title{Intrinsic constraints on very high energy emission
in gamma-ray loud blazars }
\author{Jian-Min Wang}
\affil{Laboratory of Cosmic Ray and High Energy Astrophysics,
Institute of High Energy Physics, Chinese Academy of Sciences,
Beijing 100039, and Beijing
Astrophysical Center (CAS-PKU.BAC), Beijing 100871, P.R. China,
E-mail: wangjm@astrosv1.ihep.ac.cn}
 
\begin{abstract}

Photons with very high energy up to TeV (VHE) emitted from active galactic 
nuclei (AGNs) provide some invaluable information of the origin of $\gam$-ray
emission. Although 66 blazars have been detected by {\it EGRET}, only
three low redshift X-ray selected BL Lacs (Mrk 421, Mrk 501, and 1ES 2344+514)
are conclusive TeV emitters (PKS 2155-304 is a potential TeV emitter)
since VHE photons may be absorbed by cosmological background infrared photons
({\it external} absorption). Based on the ``mirror'' effect of clouds 
in broad line region, we argue that there is an {\it intrinsic} mechanism 
for the deficiency of TeV emission in blazars. Employing the observable
quantities
we derive the pair production optical depth $\tgg(\eps_{\rm obs})$ due to 
the interaction of VHE photons with the reflected synchrotron photons by 
electron Thomson scattering in broad line region. This sets a more strong 
constraints on very high energy emission, and provides a sensitive upper 
limit of Doppler factor of the relativistic bulk motion. It has been suggested 
to distinguish the {\it intrinsic} absorption from the {\it external} by the 
observation on variation of multiwavelegenth continuum.
\end{abstract}

\section{Introduction}
Clearly, the high energy $\gam$-ray emission is an important piece
in the blazar puzzle because the $\gam$-ray observations of blazars 
provide a new probe of dense radiation field released through 
accretion onto a supermassive black 
hole in the central engine (Bregman 1990). The Energetic Gamma Ray Experiment
Telescope ({\it EGRET}) which works in the 0.1--10GeV energy 
domain has now detected 
and identified 66 extragalactic sources in 3th catalog (Mukherjee et al 1999). 
All these objects are
blazar-type AGNs whose relativistic jets are assumed to be close
to the line of sight to the observer. It seems unambiguous that the intense
gamma-ray emission is related with highly relativistic jet.

It has been generally accepted that the luminous gamma-ray emission
is radiated from inverse Compton, but the problem of seed photons remains
open for debate. The following arguments have been proposed: 
(1) synchrotron photons in jet (inhomogeneous model of synchrotron self 
Compton) (Maraschi, Ghisellini \& Celotti 1992); (2) optical and ultraviolet 
photons directly from the accretion disk (Dermer \& Schlikeiser
1993); (3) diffusive photons in broad line region (BLR) (Sikora, Begelman
\& Rees 1994, Blandford \& Levinson 1995); (4) the reflected synchrotron
photons by electron mirror in broad line region, namely, the reflected
synchrotron inverse Compton (RSC) (Ghisellini \& Madau 1996). 
These mechanisms may operate in different kinds of objects,
however there is not yet a consensus on how these mechanisms work. Also
it is not clear where the $\gam$-ray emission is taking place largely
because of uncertainties of soft radiation field in the central engine. 

On the other hand, VHE observations (Kerrick et al 1995, Chadwick 
et al 1999, Roberts et al 1999, Aharonian et al 1999) are making 
attempts to explore the radiation
mechanism because they may provide some restrictive constraints
(Begelman, Rees \& Sikora 1994, Mastichiadis \& Kirk 1997, Tavecchio, 
Maraschi \& Ghisellini 1998, Coppi \& Aharonian 1999, Harwit, Protheroe 
and Biermann 1999).
Based on the simple version of SSC model, Stecker, de Jager \& Salamon (1996)
predicted a large number of low redshift X-ray selected BL Lacs as TeV
candidates, taking into account that the presence of intergalactic 
infrared radiation field including cosmic background leads to strong 
absorption of TeV photons from cosmological emitters (Stecker 
\& de Jager 1998). It is suggested to form an extended pair halo in 
cosmological distance due to the {\it external} absorption
(Aharonian, Coppi, \& Voelk 1994). However, so far only three X-ray selected 
BL Lacs have been found to be TeV emitters by Whipple telescope ($E>300$GeV),
in addition, photons higher than 0.3TeV in the X-ray-selected PKS 2155-304 
with redshift $z=0.116$ has been detected photons 0.3TeV by Durham Mrk 6 
telescope 
(Chadwick et al 1999). The recent measurements of intergalactic infrared 
field is quite different from the previous observations (Madau et al 1998, 
Steidel 1998).  Although this {\it external} absorption is definitely 
important, the critical redshift $z_{c}$
beyond which cosmological back ground radiation and intergalactic infrared
fields will absorb VHE photons remains uncertain. 
Especially the recent VHE observations show 
that Mrk 501 emits 25 TeV photons (Aharonian et al 1999). Evidently this 
suggests that the {\it external} 
absorption can not efficiently attenuate the VHE photons from reaching us 
across distances of 100 Mpc. It is highly desired to accurately probe the 
star formation rate in order to determine the critical redshift $z_{c}$.

Thus it seems significant to study the {\it intrinsic} mechanism 
for the deficiency of TeV photons from $\gamma$-ray loud AGNs disregarding 
the absorption by intergalactic infrared radiation field. A larger Lorentz
factor of the jet implies higher density of the external photons in the
blob, if the reflection of clouds in broad line region works,
and therefore stronger absorption of high energy $\gamma$-rays
(Celotti, Fabian \& Rees 1998). Here we argue based on the hypothesis 
of Ghisellini 
\& Madau (1996) that the energy density of reflected synchrotron photon 
is high enough for pair production via interaction of gamma-ray photons 
by inverse Compton scattering with reflected synchrotron photons 
if the bulk velocity is high enough. Further we apply the present 
constraint to the representative individual objects, Mrk 421 and 3C 279. 

\section{Constraints on VHE}
Ghisellini \& Madau (1996) have calculated the energy density of reflected 
synchrotron (Rsy) emission, and compared with the other reflected components. 
They draw a conclusion that the energy density of Rsy component dominates 
over 10 times of that of reflected component of accretion disk radiation. In 
this section we make an attempt to use the observables quantities
to express the intrinsic constraints on very high energy emission.
The overall $\nu F_{\nu}$ spectrum of blazars shows that there are
two power peaks: the first is low energy one between IR/soft X-ray band,
and the second is high energy one
peaking in the MeV/GeV range (von Montigny et al 1995, Sambruna, Maraschi 
\& Urry 1996, Comastri et al 1997, Kubo et al 1998). This characteristic
can be explained by the simple context of one-zone homogeneous SSC or EC
model. The low energy peak denoted $\nu_{\rm s}$ is caused by synchrotron 
radiation of relativistic electrons, and the second peak denoted $\nu_{\rm c}$,
or $\nursc$ results from the Compton scattering off the synchrotron or
reflected synchrotron photons by the same population of electrons, 
respectively. We take the two peaks and their corresponding
fluxes as four observable quantities.

From the RSC model the magnetic field $B$ can be approximately expressed
by the observational quantities. The observed frequency of synchrotron 
photon is $\nu_{\rm s}=\cd \nu_0 \gamma_{\rm b}^2B$ ($\nu_0=2.8\times 10^6$), 
and the frequency of reflected synchrotron Compton photons reads 
$\nursc=\cd (2\Gam)^2\gamma_{\rm b}^4\nu_0B$, and we can get the 
estimation of magnetic field $B$
\begin{equation}
B=\frac{(2\Gam \nu_{\rm s})^2}{\cd \nu_0 \nursc}
 \approx \frac{\cd \nu_{\rm s}^2}{\nu_0\nursc},
\end{equation}
while in pure SSC model the magnetic field is approximately as
$B=\nu_{\rm s}^2/(\cd\nu_0\nu_{\rm c})$, 
where $\nu_{\rm c}$ is the frequency of photons emitted by SSC. The Doppler
factor $\cd=1/\Gam [1-\mu(1-\Gam^{-2})^{1/2}]$, where $\mu=\cos \theta$ is the 
cosine of the orientated angle of jet relative to the observer.
Equation (1) is similar to the model of
Sikora, Begelman \& Rees (1994) (also see Sambruna, Maraschi \& Urry 1996).
Comparing with the above two formula, we learn that RSC model needs stronger
magnetic field than SSC model does whereas the energy of relativistic 
electrons is lower in RSC model than in SSC model.
The reflected synchrotron Compton (RSC) mainly depends on two parameters:
the reflection albedo, namely, the Thomson scattering optical depth
($\tblr$), and the Lorentz factor $\Gamma$ of the relativistic jet. 

\subsection{Reflected Synchrotron Compton Emission}
In the case of power-law distribution of electrons, $N=N_0\gam^{-\al}$, 
($\gam_{\rm min}\leq \gam \leq \gam_{\rm max}$), where $N$ is the number
density of relativistic electrons, and $\gam$ is the Lorentz factor of 
electron, the synchrotron emission coefficiency is approximately given by 
$\veps_{\nu}=c_5(\al)N_0B^{1+\al \over 2}(\nu/2c_1)^{1-\al \over 2}$. 
Here $c_1=6.27\times 10^{18}$, and
$c_5(\al)$ is tabulated in Pacholczyk (1970) within the frequency range
$\nu_1\leq \nu \leq \nu_2$, where 
$\nu_{1,2}=\nu_0B(\gam^2_{\rm min},\gamma^2_{\rm max})$.
The average energy density per frequency $u^{\prime}_{\rm syn,\nu^{\prime}}$
in a region with dimension $s$
in the jet comoving frame can be obtained
\begin{equation}
u^{\prime}_{\rm syn,\nu^{\prime}}=4\pi c^{-1} c_5(\al)N_TB^{1+\al \over 2}
            \left(\frac{\nu^{\prime}}{2c_1}\right)^{1-\al \over 2},
\end{equation}
where $N_T=N_0s$. The number density of synchrotron photons can be obtained by
\begin{equation}
n^{'}_{\epsp}=n_0(\al)N_TB^{1+\al \over 2}\epsp^{-{1+\al \over 2}},
\end{equation}
and $n_0(\al)$ reads
$$
n_0(\al)=\frac{4\pi c_5(\al)}{hc}
         \left(\frac{2hc_1}{m_ec^2}\right)^{\al-1 \over 2},
$$
here $h$ is Planck constant, and $\epsp=h\nu'/m_ec^2$. 
We have employed relationship $n_{\epsp}=n_{\nu'}d\nu'/d\epsp$ to derive
equation (3).
The mean energy density is expressed by
\begin{equation}
\usynp\approx \frac{8\pi c_5(\al)}{(3-\al)c(2c_1)^{1-\al \over 2}}
              N_TB^{1+\al \over 2}\nu_2^{\prime^{3-\al \over 2}},
\end{equation}
for $\al< 3$. Defining $\lssc$ as
\begin{equation}
\lssc=\frac{L_{\rm s}}{L_{\rm ssc}}=\frac{u^{\prime}_{\rm B}}{\usynp},
\end{equation}
we have
\begin{equation}
N_T=\frac{(3-\al)c(2c_1)^{1-\al \over 2}}{64\pi^2c_5(\al)\lssc}
    B^{3-\al \over 2}{\nusp}^{\al-3 \over 2},
\end{equation}
where $\nusp$ denotes $\nu_2^{\prime}$. 

Since the opening angle of jet ($\pi/\Gam^2$) is much less than $2\pi$, 
it is then reasonable to assume that the BLR reflection approximates to 
plane mirror with thickness $\Delta R_{\rm BLR}$ and electron number
density $n_e$.
The distance distribution of reflected synchrotron photons has been
discussed by Ghisellini \& Madau (1996). The angular distribution has 
not been issued. Since the thickness of mirror
is zero, the energy density of reflected synchrotron emission 
sharply increases when blob is close to the mirror. In fact if we drop
the assumption of zero-thickness of mirror, this characteristic
will disappear. We will deal with this sophisticate model in future.
Because the reflected synchrotron emission is isotropic
in observer's frame, the blob receives the reflected photon beamed
within a solid angle $\pi/\Gam^2$. The 
subsequent section will pay attention to this effects.
Neglecting the angle-dependent distribution of reflected photon field,
we approximate the Doppler factor $\cd \approx 2\Gam$ ($\theta \approx 0$). 
For simplicity, we assume the mirror (reflecting clouds in broad
line region) has Thomson scattering optical depth 
$\tblr=\st n_e\Delta \rblr \approx \st n_e \rblr$. The received
photon density $\npp(\epspp)$ in the jet comoving frame can be then
approximately written as
\begin{eqnarray}
\npp(\epspp)&\approx &(2\Gam)^2\tblr n_0(\al)N_TB^{1+\al \over 2}
                    {\epsp}^{-{1+\al \over 2}}\nonumber \\
            & = &(2\Gam)^{3+\al}\tblr n_0(\al)N_TB^{1+\al \over 2}
                    {\epspp}^{-{1+\al\over 2}},
\end{eqnarray}
where we use $\epspp=(2\Gam)^2\epsp$ (there are two Doppler shifts due to
mirror effects). The convenient form of
cross section of photon-photon collision is (Coppi \& Blandford 1990)
\begin{equation}
\siggg=\frac{\st}{5\eps}\delta\left(\eps_0-\frac{1}{\eps}\right),
\end{equation}
where $\delta$ is the usual $\delta$-function. This approximation is
only valid for case of isotropic radiation field. The Rsy component is seen
by the blob within the solid angle 
$\Delta \Omega \approx 2\pi(1-\cos\Gam^{-1})\approx \pi/\Gam^2$. Although
the cross section of photon-photon interaction holds, the interacting
possibility among photons reduces by a factor of 
$\Delta \Omega/4\pi=1/(2\Gam)^2$ due to the beaming effects, which 
effectively reduces the opacity. Thus the pair production optical depth 
for photon with energy $\epsg$ reads
\begin{eqnarray}
\tgg^{\rm rsc}(\epsg)&=&\frac{0.2\st s}{(2\Gam)^2}
          \int \epsg^{-1}\npp(\epspp)\delta(\epspp-\epsg^{-1})
	  d\epspp \nonumber\\
          &=&(2\Gam)^{1+\al}\tblr \tgg^0(\epsg),
\end{eqnarray}
here $\tgg^0(\epsg)$ is
\begin{equation}
\tgg^0(\epsg)=0.2\st s n_0(\al)N_T B^{\al+1 \over 2}\epsg^{\al-1 \over 2}.
\end{equation}
We will show the validity of the above approximation in the next subsection.
Supposing that RSC operates efficiently
in $\gamma$-ray loud blazars, we can get $\lrsc$ from the observations
\begin{equation}
\lrsc=\frac{L_{\rm s}}{L_{\rm rsc}}
     =\frac{u^{\prime}_{\rm B}}{(2\Gam)^2\tblr \usynp}
     =\frac{\lssc}{(2\Gam)^2\tblr}.
\end{equation}
From equation (11) we have
\begin{equation}
\tblr=\frac{\lssc}{(2\Gam)^2\lrsc},
\end{equation}
and 
\begin{equation}
\tgg^{\rm rsc}(\epsg)=
     (2\Gam)^{\al-1}\left(\frac{\lssc}{\lrsc}\right)\tgg^0(\epsg).
\end{equation}
From equation (11) we know that the observed $\lrsc$ represents the
reflection ratio and Doppler factor of jet motion as long as the
Compton catastrophe does not occur. If we set $\lrsc\approx \lssc$,
we get $\tblr\approx \Gam^{-2}=0.01$ for $\Gam=10$. This value is the
lowest one in the model of Sikora, Begelman \& Rees (1994) who suggest
$\tblr=0.1\sim 0.01$. In fact we can roughly adopt $\tblr$ as the covering
factor which is usually taken to be 0.1 in fitting the broad emission 
line by photoionization model.

Inserting $B$ and $N_T$ [eqs(1) and (6)] into 
$\tgg^0$ (Eq. 10), and letting
$\epsg=\eps_{\rm obs}/\cd$  and $s=c\cd \dt$, 
we have the pair production optical depth for $\eps_{\rm obs}$
in the observer's frame
\begin{eqnarray}
\tggrsc& =&K_{\alpha}\frac{\nu_{\rm s}^{5+\al \over 2}}{\nu^2_{\rm rsc}}
 (2\Gam)^{3+\al} \cd^{1-\al} \eps_{\rm obs}^{\al-1 \over 2}
 l^{-1}_{\rm rsc}\dt\nonumber\\
   &\approx& K_{\alpha}\frac{\nu_{\rm s}^{5+\al \over 2}}{\nu^2_{\rm rsc}}
             \cd^4 \eps_{\rm obs}^{\al-1 \over 2} \lrsc^{-1}\dt,
\end{eqnarray}
where $K_{\alpha}=\frac{(3-\al)\st c}{80\pi h \nu_0^2}
             \left(\frac{h}{m_ec^2}\right)^{\al-1 \over 2}$
($K_{\alpha}=7.9\times 10^{-18}$ for $\alpha=2.4$).
There are five observational parameters: $\nu_{\rm s}$, $\nu_{\rm rsc}$,
$\al$, $\dt$ and $\lrsc$; and the unknown Doppler factor $\cd$.
For the typical value of parameters, $\al=2.4$, 
$\nu_{\rm s}=4.0\times 10^{14}$Hz, and $\nu_{\rm rsc}=1.0\times 10^{25}$Hz,
$\dt=1$~day, and $\cd=10$, we have
\begin{eqnarray}
\tggrsc&=&1.9~l^{-1}_{\rm rsc}\cd_{10}^4
                     \left(\frac{\eps_{\rm obs}}{{\rm TeV}}\right)^{0.7}
                     \left(\frac{\dt}{{\rm day}}\right)\nonumber\\
            & &\left(\frac{\nu_{\rm s}}{4.0\times 10^{14}{\rm Hz}}\right)^{3.7}
            \left(\frac{\nu_{\rm rsc}}{10^{25}{\rm Hz}}\right)^{-2},
\end{eqnarray}
where $\cd_{10}=\cd/10$. Figure 1 shows the opacity due to pair production
of photons with very high energy encountering with the reflected synchrotron
photons. The equation (15) tells us the constraints on VHE from jet:
(1) smaller $\lrsc$, i.e. stronger reflection, will leads to the absorption 
of TeV photons. This parameter represents the energy density reflected by 
the BLR cloud including the bulk relativistic motion. 
From this estimation we know that TeV photon will be absorbed by the
reflected synchrotron photons provided that $\lrsc<1.9$.
(2) $\tgg$ is sensitive to $\nu_{\rm s}$ and $\nu_{\rm rsc}$.
(3) $\tgg$ is proportional to $\cd^4$, in contrast to the usual down-limit 
(see Mattox et al 1993, and Dondi \& Ghisellini 1995), providing the upper 
limit Doppler factor of bulk motion  from $\tgg\leq 1$,  
\begin{eqnarray}
\cd &\leq &8.5~l^{1/4}_{\rm rsc}
     \left(\frac{\eps_{\rm obs}}{{\rm TeV}}\right)^{-0.175}
     \left(\frac{\dt}{{\rm day}}\right)^{-0.25}\nonumber\\
    & &\left(\frac{\nu_{\rm s}}{4.0\times10^{14}{\rm Hz}}\right)^{-0.925}
       \left(\frac{\nu_{\rm rsc}}{10^{25}{\rm Hz}}\right)^{0.5}.
\end{eqnarray}
This is a new constraint, which is expressed by the observational quantities.
It lends us a simple and efficient way to select TeV candidates from known
blazars in term of their known characteristics. 

\subsection{Angular Distribution of Reflected Photons}
The received photons reflected by BLR in comoving frame is anisotropic, 
therefore, the pair opacity should be carefully treated.
We have made important approximations that the BLR is thought to be
a plane mirror and treated the photon-photon interaction in an
approximate way. Now let us show the validity of this approximation.
We adopt the geometry shown in Fig 1c of Ghisellin \& Madau (1996).
They show that the energy density of reflected synchrotron
photon strongly depends on the location of emitting blob.
We should admit that the aximal symmetry holds in the reflected synchrotron
emission. We approximate the Thomson scattering event by
isotropic scattering with cross section $\st$ and neglect
recoil, which is a very good approximation when $\eps_s \ll 1$.
The angular distribution of reflected synchrotron emission
is given by $n_{\rm ph}(\eps_s, \mu, r_0)=n_0 f(\mu,r_0)\eps_s^{-q}$ ($n_0$
is a constant), the
function $f(\mu, r_0)$ determines the angular distribution of reflected
synchrotron photons (Ghisellini \& Madau 1996)
\begin{equation}
f(\mu,r_0)=\cd^2(\mu)g(\mu,r_0),
\end{equation}
where 
$g(\mu,r_0)=\left[\left(1-r_0^2+r_0^2\mu^2\right)^{1/2}-r_0\mu\right]^{-2}$,
and $r_0=R_{\gam}/\rblr \in (0,1)$, where $R_{\gam}$ is the distance of
blob to the center. Figure 2 shows the angular distribution in blob comoving 
frame. It can be seen that the geometry effect of reflecting mirror isotropizes
the radiation at some degrees, but the beaming effect still dominates. It is
still a good approximation that the radiation is beamed with a cone of
solid angle $\pi/\Gam^2$. 
Thus the pair opacity can be written as (Gould \& Schr\'eder 1967)
\begin{equation}
\tgg(\epsg)=2\pi\rblr \int_0^1 dr_0 \int_{-1}^1d\mu (1-\mu)\int_{\eps_c}
     n_{\rm ph}\siggg d \eps_s,
\end{equation}
where $\eps_c=2/(1-\mu)\epsg$ and the photon-photon cross section
$\siggg$ reads
\begin{equation}
\siggg =\frac{3\st}{16}(1-\beta^2)\left[(3-\beta^4)\ln\left(
       \frac{1+\beta}{1-\beta}\right)-2\beta (2-\beta^2)\right],
\end{equation}
where $\beta$ is the speed of the electron and positron in the center of
momentum frame $\beta=\left[1-2/\epsg \eps_s(1-\mu)\right]^{1/2}$.
Performing the integral we have
\begin{equation}
\tgg=n_0\st\rblr \epsg^{q-1}A(q),
\end{equation}
where $A(q)$ reads
\begin{equation}
A(q)=2^{3-q}\pi A_0(q)\int_{-1}^1d\mu(1-\mu)^qA_1(\mu),
\end{equation}
with
\begin{equation}
A_0(q)=\int_0^1d\beta (1-\beta^2)^{q-2}\beta \siggg(\beta),
\end{equation}
and $A_1(\mu)=\int_0^1f(\mu,r_0)dr_0$ is the integral of $f(\mu,r_0)$ over 
the entire broad line region, which can be evaluated as
\begin{eqnarray}
A_1&=&\sqrt{1-\mu^2}\nonumber\\
   & &\left\{\cos \theta_0 \ln \left[
      \frac{\tan {\phi_1 \over 2}}{\tan {\phi_2 \over 2}}\right]
      +\sin \theta_0 \ln \left[\frac{\sin \phi_1}{\sin \phi_2}\right]
                                                          \right\},
\end{eqnarray}
with $\phi_1=\pi/2-\theta_0$, $\phi_2=\arccos \sqrt{1-\mu^2}-\theta_0$,
and $\theta_0=\arcsin \sqrt{1-\mu^2}$.
The function $A(q)$ is plotted in Fig 3.  Since the beamed radiation
field reduces the effective cross section of photon-photon interaction
by a factor $1/(2\Gam)^2$, it would be convenient to check our
approximation by the quantity $(2\Gam)^2A(q)$. We can easily find that it is
close to 0.3$\sim$0.4 when $q \sim 1.7$, suggesting our approximation is 
accurate enough. It should be pointed out that the present treatments of
reflected synchrotron radiation can be conveniently extended to the inclusion
of the radiation from the secondary electrons if we further study the pair 
cascade in the future. 

\subsection{The Dimension of External Absorption}
The last two subsections are devoted to the {\it internal} absorption of TeV
photons, the developments of pair cascade due to the present mechanism
will be treated in a preparing paper (Wang, Zhou \& Cheng 2000).
However it would be useful to compare the dimensions and radiation of
the pair cloud due to the {\it internal} absorption and the pair halo
suggested by Aharonian, Coppi, \& Veolk (1994), who argue the formation of 
pair halo due to the interaction of TeV photons from AGNs with infrared
photons of cosmological background radiation. This {\it external}
absorption produces pairs, which are quickly isotropized by an ambient
random magnetic field, forming a extended halo of pairs with typical 
dimension of ($R>1$Mpc). Without specific mechanism we know that 
the time scale of halo formation is of about $10^6$ yr. Usually this 
absorption is regarded as the main mechanism of deficiency of TeV 
emission from {\it EGRET}-loud blazars (Stecker \& de Jager 1998). 
Let us simply estimate the scale of pair halo before it is isotropized 
by the ambient magnetic field. Assuming the intergalactic magnetic
field $B=10^{-9}$ Gauss, then the mean free path of pair electrons in
halo reads
\begin{equation}
\lambda_{\rm e}\approx 1.0 \left(\frac{E}{1.0{\rm TeV}}\right)^{0.5}
            \left(\frac{B}{10^{-9}{\rm G}}\right)^{-0.5} ~{\rm Kpc},
\end{equation}
The initial halo is of such a dimension, which is much larger than
that of {\rm intrinsic} absorption case. Aharonian, Coppi, \& Volk
(1994) have suggested some signatures of such an extended halo,
especially for the light curves in high energy bands (Coppi \&
Aharonian 1999).
Anyway this is much larger than that of the present {\rm internal} 
pair cloud. Thus it is easier to distinguish the two cases.

\section{Applications}
We have set a new constraint on the very high energy emission in term
of observable quantities. As the applications of the present model, 
we would like to address some properties of very high energy from blazars.

\subsection{Broadband Continuum and Mirror}
The broadband continuum of blazars show attractive features which indicate
the different processes powering the objects.
The ratio $L_{\gam}/L_{\rm op}$
of $\gam$-ray luminosity to optical in flat spectrum radio quasars (FSRQs)
is quite different from that in BL Lacs (Dondi \& Ghisellini 1995).
Comastri et al (1997) confirmed this result in a more larger samples
and found this mean ratio
is roughly of unity in BL Lacs and $L_{\gam}/L_{\rm op}\approx 30$,
namely $l_{\rm rsc}\approx 0.03$ in FSRQs. 
Ghisellini et al (1993), using the classical limit of SSC model,
show that there is a systematical difference in Doppler factors $\cd$ between
BL Lacs and core-dominated quasars, $\langle \log \cd \rangle=0.12$ for 
BL Lacs and $\langle \log \cd \rangle=0.74$ for core-dominated quasars. 
These differences have been confirmed by G\"uijosa \& Daly (1996) who assume
that the particles and magnetic field are in equipartition.
This difference would lead to more prominent difference of reflected 
synchrotron photon energy density, suggesting a different mechanism in 
these objects. The two systematically different features in
$\lrsc$ and Doppler factor $\cd$ strongly suggest that the different 
mechanism of $\gam$-ray 
radiation may operate in these objects. From eq.(15) it is believed that
the deficiency of TeV emission in radio-loud quasars may be {\it intrinsic}
due to the present mechanism. 

\subsection{On Mrk 421 and 3C 279}
Whipple observatory had ever searched for TeV gamma-ray emission for 15
EGRET-AGNs with low redshift, but only Mrk 421 has positive signal(Kerrick et
al 1995). Even at present stage only three X-ray selected BL Lacs have been
reported as TeV emitters, Mrk 421, Mrk 501 and
1ES2344+514 (Cataness et al 1997), and PKS 2155-304 is a potential TeV
emittor (Chadwick et al 1999).
We can apply the present model to the two representative sources:
Mrk 421 and 3C 279 for specific illustration.

{\it Mrk 421:} This is an X-ray selected BL Lac object, and has been
detected GeV $\gam$-ray emission by EGRET (Lin et al 1992), and the first
TeV emission by Whipple (Punch et al 1992). It has been extensively and
frequently observed by telescopes from radio to TeV bands(Kerrick et al 1995,
Macomb et al 1995, Takahashi et al 1996, Krennrich et al 1999).
TeV observations of Mrk 421 by Whipple show that the TeV photon did
not flare much more dramatically than the X-rays, suggesting that the 
enhanced high-energy electrons were scattering off a part of the 
synchrotron spectral energy distribution that remained constant 
(Takahashi et al 1996). Roughly
speaking this object satisfies the energy equipartition for the two
power peaks(Zdziarski \& Krolik 1993, Macomb et al 1995), suggesting
$\lssc\approx 1$ (Macomb et al 1995), and pure SSC model agrees with the 
observations (Krennrich et al 1999), suggesting $\lrsc\gg 1$. This 
indicates the RSC process is not important. The
synchrotron component peaks in luminosity at UV to soft X-ray energies
and continues into KeV X-rays(Maraschi, Ghisellini \& Celotti 1994).
The gamma-ray emission extends from 50 MeV to an astounding TeV. Data
combined over several periods (Lin et al 1992) reveal a hard GeV
spectrum ($\al_{\rm GeV}\approx 0.7$) by EGRET and a steeper one at TeV
energies ($\al_{\rm TeV}\approx 1.30$) from the Whipple observatory
(Schubnell et al 1994), implying a spectral break. 
The multiwavelegenth spectrum shows that
$\nu_{\rm s}=3\times 10^{16}$Hz, $\nu_{\rm rsc}=6.5\times 10^{25}$Hz,
and $\alpha=2.0$ (Macomb et al 1995, Kubo et al 1998). The  shortest
timescale of $\gam$-ray variability is about $\dt \approx 20$ minutes 
(Gaidos et al 1996). If we take $\cd=10$ and $\lrsc=1$, we have 
$\tggrsc=5.0\times 10^5$ for TeV photons.  This means $\lrsc \gg 1$,
which is consistent with pure SSC model. From this we can estimate the 
scattering medium $\tblr=2.0\times 10^{-4}$ [$\tggrsc=1$]
when we take $\lrsc=5.\times 10^{-5}$ and $2\Gam=10$,
suggesting the mirror effects can be ruled out in this object.
This result agrees to the absence of any evident emission lines in Mrk 421.
Interestingly, Celotti, Fabian \& Rees (1998) have suggested from rapid 
TeV variability of Mrk 421
that its accretion rate is lower than $10^{-2}\sim 10^{-3}$ Eddington rate.
They thus propose that advection-dominated accretion flow or ion pressure 
supported tori (Ichimaru 1977, Rees et al 1982, Narayan \& Yi 1994) may 
power the luminosity. 

{\it 3C 279:} This is a typical FSRQ. The first Whipple observation shows 
negative signal in TeV
(Kerrick et al 1995), however the multiwavelength simultaneous
observations by {\it EGRET}, {\it ASCA}, {\it RXTE}, {\it ROSAT}, {\it IUE}
 in 1996 January-February
show an intensive flare with very flat spectrum in {\it EGRET} band (Wehrle
et al 1998), showing $\lrsc\approx 10^{-1}$ 
at the high state of $\gam$-ray emission. It has been argued that
RSC may explain the 1996 gamma-ray flare (Wehrle et al 1998). 
The observed flare show that
the synchrotron emission peaks at $\nu_{\rm s}=5.0\times 10^{12}$Hz,
$\nu_{\rm rsc}=1.0\times 10^{23}$Hz, $\lrsc=0.1$, and $\dt=$8hr
(Wehrle et al 1998). If we take 
$\cd=10$, then we have $\tggrsc=3.2$. Therefore it is expected that
no TeV emission occurs in this object. However it is interesting to note
that this is due to the {\it intrinsic} mechanism. We hope that
there will be some effects due to the presence of pair production in 
the VHE flare (Wang, Zhou \& Cheng 2000). 
The Q1633+382 (Mattox et al 1993) is quite similar to 3C 279, but it
shows much smaller $\lrsc < 10^{-2}$. The strong reflected synchrotron
photons as seed photons may appear in this object, however its redshift
($z=0.181$)
is too large to detect VHE photons due to the absorptions of back ground 
photons. Here we suggest that the deficiency of VHE emission may be
{\it intrinsic}. It is expected to make simultaneous
observations at other bands to test its light curves in order to 
reach a decision.

\section{Conclusions and discussions}

The present paper focuses our attention on the effects of BLR mirror on the
attenuation of $\gam$-ray in blazars. The mirror effect mainly depends 
on two parameters: Lorentz factor of the 
bulk motion ($\Gamma$) and the Thomson scattering depth ($\tblr$) of 
broad line region.
Based on the calculations, we would like to draw the conclusions:

1) The parameters $\lrsc$ and Doppler factor $\cd$ in FSRQs are systematically 
greater than that in BL Lacs. This will cause the more stronger 
``{\it intrinsic}'' absorption of VHE photons in FSRQs than that in BL Lacs.
It is predicted that there is general absence of very high energy emission
in FSRQs, owning to the attenuation of VHE photons by the BLR reflection of
synchrotron emission. 

2) The mirror model provides a new constraint on relativistic bulk motion.
That {\it intrinsic} absorption of TeV photons may operate in
some objects, especially in FSRQs. This constraint is cause by the motion
of blob itself.

Although the origin of $\gam$-ray emission in blazars still remains open,
VHE observations sets strong constraints on blazar's radiation mechanism. 
These constraints are:
(1) brightness temperature exceeding the Kellermann-Pauliny-Toth (Begelman 
et al 1994), (2) multiwavelength light curves based on the homogeneous
model (Mastichiadis \& Kirk 1997), (3) high energy variations in X-ray
and $\gam$-ray including interaction with background IR radiation (Coppi
\& Aharonian 1999). These constraints are mainly based on SSC model. 
The deficiency of VHE photons from high redshift $\gamma$-ray loud blazars
may be explained by the interaction of the cosmic background radiation
fields with the VHE photons. However the possible alternative mechanism
may be due to the {\it intrinsic} attenuation by the reflected synchrotron
photons. Three BL Lac objects have been found to show $H\alpha$ and $H\beta$
emission, indicating the existence of broad line region in these objects
(Vermeulen et al 1995, Corbett et al 1996), even Mrk 421 has been detected 
weak luminosity of a broad emission line (Morgani, Ulrich \& Tadhunter 1992). 
The increasing evidence of the presence of broad emission lines
in BL Lacs lends the possibility that the reflected external photons might
be the main source of seed photon in this kind of blazars. Distinguishing 
the two different mechanisms might be traced by 
the following-up observations in other wavebands because a pair 
cascade process may be developed, forming a pair halos in the {\it external}
absorption (Aharonian, Coppi \& Voelk 1994).
Such an extended halo due to {\it external} absorption may be of very 
long variable timescale at least $\sim 10^3$ yr (corresponding to one
mean free path) [see equation (24)]. 
However if  the {\it intrinsic} absorption works in the central
engine, the time-dependent synchrotron self-Compton model including
pair cascade (Wang, Zhou \& Cheng 2000) could predict
the interesting spectrum and light curves, which may interpret the variations
of PKS 2155-304 (Urry et al 1997). 
The other radiative properties of such an extended pair halo are needed to
be studied in order to distinguish the {\it intrinsic} absorption
from the {\it external} one.

\acknowledgements{
The author is very grateful to the anonymous referee for the physical
insight of comments and suggestions, especially on the discussions on
the angular distribution of reflected synchrotron photons and its
effects on the opacity of pair production. I thank C.-C. Wang and F.-J. Lu
for their careful reading of the manuscript and interesting discussions.
The simulating discussions with Y.-Y. Zhou, T.-P. Li, M. Wu and 
B.-F. Liu are gratefully acknowledged. This research is supported by 
'Hundred Talents Program of CAS' and Natural Science
Foundation of China under Grant No. 19803002.}

\newpage
\begin{figure}
\vspace{7.5cm}
\epsscale{9.0}
\plotfiddle{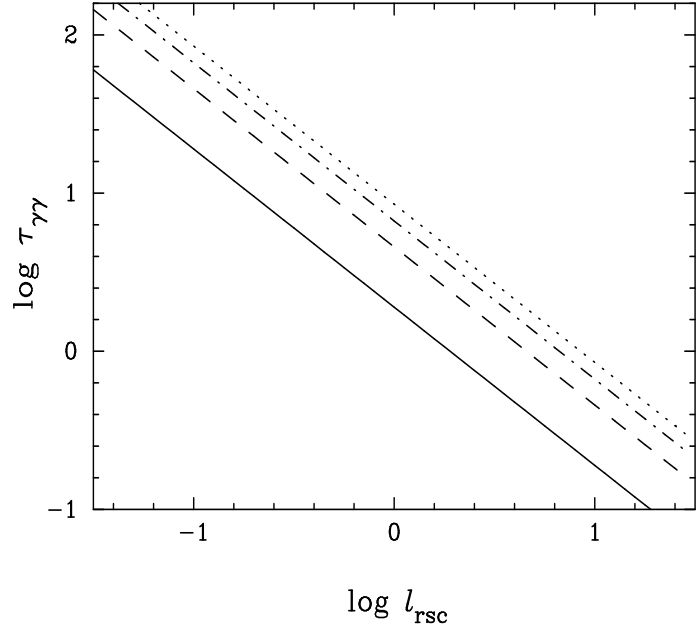}{190pt}{-90}{75}{85}{-295}{560}
\vspace{-30mm}
\caption{The opacity of pair production vs the ratio of two peak fluxes
in the continuum of blazars. We set $\cd=10$, $\dt=1$day, 
$\nu_{\rm s}=4.0\times 10^{14}$Hz, and $\nursc=10^{25}$Hz.
The lines from the below to upper
are corresponding to the opacity of very high energy photon with 
$\eps_{\rm obs}=1.0,3.5,6.0,8.5$TeV, respectively.}
\label{fig1}
\end{figure} 

\newpage
\begin{figure}
\vspace{7.5cm}
\epsscale{9.0}
\plotfiddle{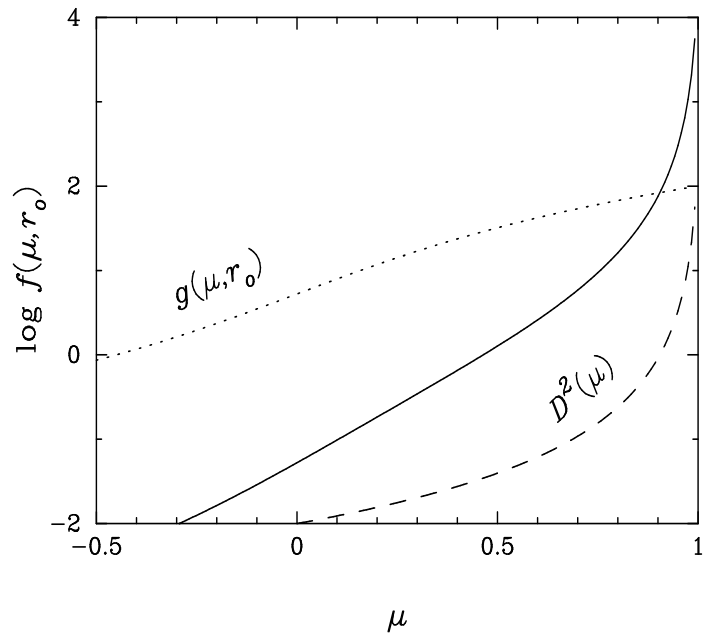}{190pt}{-90}{75}{85}{-295}{560}
\vspace{-30mm}
\caption{The angular distribution of reflected synchrotron photons.
The location is $r_0=0.9$. The solid line represents the function
of angular distribution $f(\mu,r_0)$.}
\label{fig2}
\end{figure} 

\newpage
\begin{figure}
\vspace{7.5cm}
\epsscale{9.0}
\plotfiddle{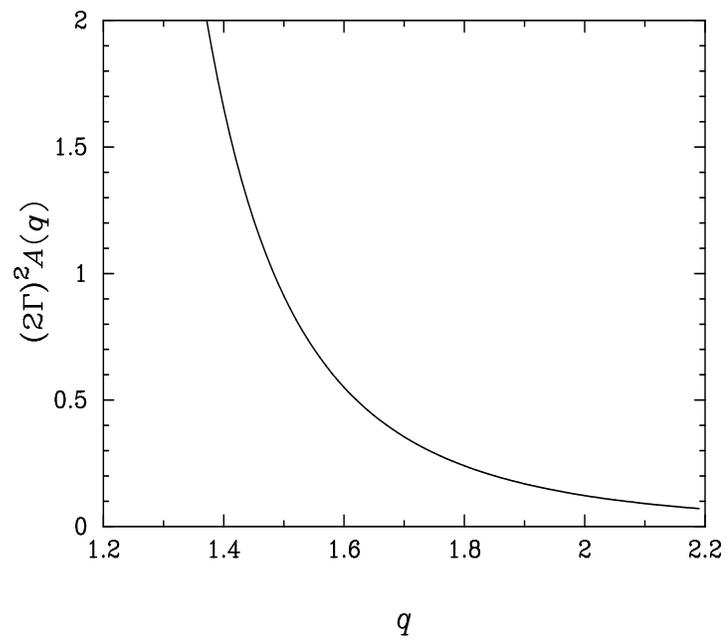}{190pt}{-90}{75}{85}{-295}{560}
\vspace{-30mm}
\caption{The plot of $A(q)$ vs $q$ (eq. 21). We multiply $A(q)$ by
the factor $(2\Gam)^2$ in order to see the reduced cross section of
photon-photon interaction in anisotropic radiation field. We take
$\Gam=10$.}
\label{fig3}
\end{figure} 

\end{document}